\documentstyle[epsf,astrcite]{mn}
\newcommand{\un}[2]{\mbox{\rm\thinspace #1$^{#2}$}}

\newcommand{\be}[1]{\begin{equation}\label{#1}}
\newcommand{\ee}{\end{equation}}

\newcommand{\degs}{\mbox{$^{\circ}$}}

\newcommand{\etal}{et~al.}
\newcommand{\Fig}[1]{Fig.\,\ref{#1}}

\newcommand{\gsim}{\mathrel{\hbox{\rlap{\lower.55ex \hbox {$\sim$}}
                   \kern-.3em \raise.4ex \hbox{$>$}}}}
\newcommand{\lsim}{\mathrel{\hbox{\rlap{\lower.55ex \hbox {$\sim$}}
                   \kern-.3em \raise.4ex \hbox{$<$}}}}
\newcommand{\msun}{\mbox{M$_\odot$}}

\newcommand{\Sect}[1]{Sect.\,\ref{#1}}
\newcommand{\sect}[1]{sect.\,\ref{#1}}

\newcommand{\sub}[1]{_{\rm #1}}

\newlength{\ffh}

\newcommand{\muc}{\mbox{$\mu\sub{c}$}}
\newcommand{\me}{\mbox{$\dot{M}_{-8}$}}
\title[Accretion-induced field decay]{
       Evidence against field decay proportional to accreted mass in 
       neutron stars}
\author[Ralph Wijers]{Ralph A.M.J. Wijers\\
        Institute of Astronomy, Madingley Road, Cambridge CB3 0HA, UK\\
	Email: R.Wijers@ast.cam.ac.uk}
\date{submitted to MNRAS, 24-7-96, revised 27-11-96, accepted 02-01-97} 

\begin{document}
\maketitle
\begin{abstract}
A specific class of pulsar recycling model, in which magnetic-field
decrease is a function only of the amount of mass accreted onto the
neutron star, is examined in detail. It is shown that no model in this
class is consistent with all available data on X-ray binaries and recycled
pulsars.  Only if all constraints are stretched to their limit and a few
objects (PSR\,B1831$-$00 and 4U\,1626$-$67) are assumed to
have formed in a non-standard manner is there still an acceptable model of
this kind left.  Improved measurements of the parameters of a few of the
oldest known radio pulsars will soon test and probably rule out that one
as well.  Evidence for the origin of PSR\,B1831$-$00 via accretion-induced
collapse of a white dwarf is called into question as a result.
\end{abstract}
\begin{keywords}
pulsars: recycled -- neutron stars: structure --
          neutron stars: magnetic fields -- X-ray binaries
\end{keywords}

   \section{Introduction}
   \label{intro}

The growing body of evidence against rapid spontaneous magnetic-field decay
in neutron stars (Kulkarni 1986; Verbunt, Wijers, \& Burm 1990; 
Bhattacharya et~al.\ 1992)\nocite{kulka:86,vwb:90,bwhv:92} has led to
the hypothesis that pulsar magnetic fields are lowered by recycling in
binaries (see Bhattacharya \& Van den Heuvel 1991\nocite{bh:91} for a
review).  The mechanism for this remains unclear, with suggestions
including decay of crustal fields due to heating (Blondin \& Freese 
1986)\nocite{bf:86} or burial of
the field (Bisnovatyi-Kogan \& Komberg 1975; Romani 1990, 1993)
\nocite{bk:75,roman:90,roman:93} and decay of core fields due to
flux tube expulsion from the superfluid interior (Srinivasan et~al.\ 
1990)\nocite{sbmt:90}.

A heuristic model first suggested by Taam and Van den Heuvel 
(1986)\nocite{th:86}
and explored further by Shibazaki \etal\ (1989)\nocite{smsn:89} is that the
ratio of initial to final magnetic field is simply proportional to the
amount of mass accreted. The detailed model by Romani (1993)\nocite{roman:93}
supports this in the regime of high accretion rates. Van den Heuvel and
Bitzaraki (1995)\nocite{hb:95} have shown that it is consistent with the
magnetic fields of most recycled pulsars with helium white dwarf companions.
In this paper the model (and a generalised version of it) is tested for
agreement with a wider body of data including many types of neutron star
binary, and it fails the test.

The details
of the model used to evolve pulsar periods and fields is discussed in 
Section~\ref{model}. The constraints derived from applying the model to
observed neutron stars are presented in Section~\ref{bdmcon};
some further discussion of the implications (Section~\ref{discu}) and
an outline of the main conclusions (Section~\ref{conclu}) follow.

   \section{The model}
   \label{model}

      \subsection{Neutron star structure}
      \label{neustruc}

The amount of mass accreted onto recycled pulsars can approach \msun,
so we should follow the concomitant change of radius and moment of inertia.
To compute the gravitational mass, $M\sub{g}$, as a function of
baryon mass, $M\sub{b}$, I use the binding energy correction
given by Lattimer \& Yahil (1989)\nocite{ly:89}:
\be{eq:ebind}
   M\sub{b} = M\sub{g} + 0.084\msun \left(\frac{M\sub{g}}{\msun}\right)^2
\ee
As $M\sub{b}$ increases by 
accretion, $M\sub{g}$ increases more slowly. The mass dependence of the radius
is taken to be
\be{eq:rad}
   R_6 = \frac{R}{10\un{km}{}} = \left(\frac{M\sub{g}}{1.4\msun}\right)^{-1/3}
\ee
and assuming the radius of gyration is a constant fraction of $R$ the
moment of inertia will be
\be{eq:I}
   I_{45} = \frac{I}{10^{45}\un{g}{}\un{cm}{2}} = 
       \left(\frac{M\sub{g}}{1.4\msun}\right)^{+1/3}.
\ee
Here the normalisation is chosen to get the `standard' values $I_{45}=1$
and $R_6=1$ for a $1.4\msun$ neutron star. Test calculations show that 
modest deviations from the above do not affect the conclusions. After
accreting $1\msun$ onto a $1.4\msun$ neutron star
we have $M\sub{g}=2.05\msun$, which 
exceeds the maximum mass for a significant number of equations of state.
To be conservative, I shall just assume that accretion does not lead
to the collapse of the neutron star into a black hole rather than
use the maximum mass as an extra constraint.

      \subsection{Magnetic field evolution}
      \label{bevol}

Since both the magnetic fields and the amounts of accreted mass of neutron
stars vary by many powers of ten, any model relating the two will almost
necessarily be a power law:
\be{eq:bm}
   \frac{B}{B_0} = \left(1+\frac{\Delta M}{M\sub{c}}\right)^{-\beta}.
\ee
$B$ and $B_0$ are the current and initial field, $\Delta M$ is the
accreted (baryon) mass, and $M\sub{c}$ is the characteristic mass at which
field decrease becomes significant. For $\beta=1$ this is the well-known
scaling relation proposed by Taam \& Van den Heuvel (1986)\nocite{th:86} and
Shibazaki \etal\ (1989)\nocite{smsn:89}. I have added the exponent because the
generalisation probably encompasses effectively all models in which $B$
depends on $\Delta M$ only.  Once the amount of accreted mass becomes
enough to affect the radius of the neutron star, one should perhaps
consider changes in the surface field due to conservation of magnetic flux
$BR^2$, but here I use Eq.~\ref{eq:bm} throughout.

      \subsection{Spin evolution}
      \label{Pevol}

Another constraint on the decay model is imposed by the fact that
some recycled pulsars are spinning very rapidly, near their equilibrium
period.  For low fields and/or low accretion rates the accreted mass cannot
carry the required angular momentum to the neutron star.  To follow the
evolution in this phase I use the torque model for disc accretion by Ghosh
\& Lamb (1979)\nocite{gl2:79}.  The spin-up rate is given by
\be{eq:GLpdot}
   \dot{P} = -3.6\times 10^{-4} B_{12}^{2/7} R_6^{6/7}I_{45}^{-1}M_{1.4}^{3/7}
              \dot{M}_{-8}^{6/7} P^2 n(\omega\sub{s})~~~\un{s}{}\un{yr}{-1},
\ee      
where $B_{12}= B/10^{12}\un{G}{}$ and 
$\dot{M}_{-8}=\dot{M}/10^{-8}\un{\msun}{}\un{yr}{-1}$.
 $\omega\sub{s}$ (called fastness) is
the ratio of neutron star spin frequency to the Kepler frequency of an
orbit just outside the magnetopause and $n(\omega\sub{s})$ is the
dimensionless torque function which is given approximately by
\be{eq:noms}
   n(\omega\sub{s}) = 1.39
   \frac{1-\omega\sub{s}[4.03(1-\omega\sub{s})^{0.173}-0.878]}{
   1-\omega\sub{s}}.
\ee
The fastness is related to the spin period by
\be{eq:oms}
   \omega\sub{s} = 0.43 B_{12}^{6/7}R_6^{18/7}M_{1.4}^{-5/7}
                   \dot{M}_{-8}^{-3/7} P^{-1}.
\ee   
($M_{1.4}=M\sub{g}/1.4\msun$)
Since the physics of the interaction between accretion disc and
magnetosphere is uncertain and the expressions also depend on
the poorly understood
structure of the accretion disc itself, there is ample uncertainty in the
torque. For the present purpose it is adequate, however. The value of the
critical fastness, $\omega\sub{cr}$, at which the torque vanishes (0.349 for
Eq.~\ref{eq:noms}) is important, so the effect of changing it is examined
below. Setting $\omega\sub{s}=\omega\sub{cr}$ in Eq.~\ref{eq:oms} fixes a
relation $P\sub{eq}\propto B^{6/7}$ called the spinup line or equilibrium
spin period.

      \subsection{Initial conditions}
      \label{initcon}

In the calculations all neutron stars start with a gravitational mass,
$M\sub{g}$, of $1.4\msun$ (so $M\sub{b}=1.56\msun$). The accretion rate is
taken to be constant, with a usual value of
$2\times10^{-8}\un{\msun}{}\un{yr}{-1}$, the Eddington rate for electron
scattering opacity on material with hydrogen mass fraction $X=0.7$ onto a
10\,km neutron star. Pulsars with initially high fields retain no memory
of their initial spin periods once they reach fields below about $10^{10}$\,G
when they are evolved using the above equations. Since the only pulsars to
which spin constraints are applied here have lower fields than this,
the initial period does not matter, and all pulsars are started at
$P=P\sub{eq}$. For very short-lived accretion phases that start
when the pulsar is still spinning very rapidly (e.g.\ Brown 
1995\nocite{brown:95})
this is not correct, because the pulsar will then eject mass thrown towards
it and never accrete. This more complicated situation is treated elsewhere
(Wijers, Braun, \& Brown 1997\nocite{wbb:97}).

The initial magnetic field is that of young radio pulsars, which at birth
have $\log B_0=12.3$ typically, with a spread around that of a factor 2.
To investigate the influence of the initial field I compute all results
for $\log B_0=$ 11.8, 12.3, and 12.8.  This range nominally covers 87\% of
all pulsars in the best-fit initial magnetic field distribution found by Hartman
et~al.\ (1996)\nocite{hbwv:96}. In assembling the final constraints on
the parameters $M\sub{c}$ and $\beta$ I use whichever value in this range
is least constraining to the accretion-induced decay model. Therefore all the
derived bounds are necessarily somewhat statistical in nature rather than
absolute, because some 7\% of pulsars will lie outside an indicated range
on either side.

      \subsection{Example evolution tracks}
      \label{examp}

\begin{figure}
   \epsfxsize=84mm\epsfbox{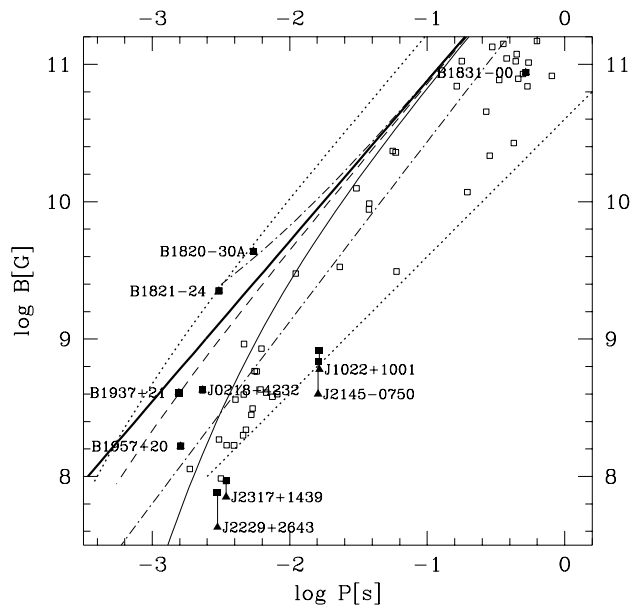}
   \caption{The lower left corner of the $B$-$P$ diagram of radio pulsars.
	    Open squares denote pulsars, filled symbols with names are those
	    specifically discussed in this paper. The lower dotted line 
	    indicates a characteristic age, $P/2\dot{P}$, of 10\,Gyr,
	    the heavy solid line
	    is the standard spinup line. The other curves
	    are evolution tracks of pulsars, explained in the text.}
   \label{fi:bp}
\end{figure}

Fig.~\ref{fi:bp} shows the lower left corner of the radio pulsar
period-magnetic field diagram with the known pulsars from the Princeton
catalogue (Taylor, Manchester, \& Lyne 1993\nocite{tml:93}, as updated by
Taylor \etal\ 1995\nocite{tmlc:95}) indicated as open symbols. The filled
symbols with names are the pulsars that are discussed below. Their
properties are listed in the top part of Table~\ref{ta:systems}. For
pulsars close to the Hubble line (lower dotted curve) the value of the
magnetic field derived from the period derivative ($B\sub{m}$ in
Table~\ref{ta:systems} and filled squares in \Fig{fi:bp}) should be
corrected for the proper motion of the pulsar which contributes to the
measured period derivative (Shklovskii 1970; Camilo, Thorsett, \& Kulkarni
1994)\nocite{shklo:70,ctk:94}. The corrected values ($B\sub{e}$ in
Table~\ref{ta:systems} and filled triangles in Fig.~\ref{fi:bp}) are
estimates based on an assumed 75\un{km}{}\un{s}{-1}\ transverse velocity,
except for PSR\,J2317+1439, where a transverse velocity of
70\un{km}{}\un{s}{-1}\ has been measured (Camilo, Nice, \& Taylor
1996a)\nocite{cnt:96}.

All evolution tracks shown in Fig.~\ref{fi:bp} are for $\log B_0=12.3$ and
$\me=2$.
The dashed-dotted curves are for $\beta=1$ and
$\muc\equiv\log(M\sub{c}/\msun)=-5$ (lower) and $\muc=-3$ (upper).
The curve for low \muc\ passes nicely
through the millisecond pulsar range but misses the fastest ones. For high
\muc\ the lowest fields are not reached (the tracks in Fig.~\ref{fi:bp}
all end at $\Delta M=0.8\msun$). The intermediate case $\muc=-4$ (not
shown) almost follows the standard spinup line (heavy solid curve) and is
acceptable. The curve for $\muc=-3$ illustrates the importance of the
changing neutron star mass:  PSR\,B1821$-$24 appears to be above the
spinup line, but that is drawn for a fixed neutron star mass of
$1.4\msun$, as is traditional. Increasing the mass shifts the spinup line to
the left enough to reach this pulsar without exceeding $\me=2$.

Low fields can be reached with $\muc=-3$ by increasing $\beta$. This is
illustrated by the dashed ($\beta=1.5$) and solid ($\beta=2$) curves.
The dashed curve is fully acceptable for any of the fast, low-field pulsars,
whereas the $\beta=2$ case does not produce the fast ones because the field
has decayed to low values before enough mass has been accreted to spin the
pulsar up. To shift the evolution track further to the left so that 
 PSR\,B1821$-$24 can be reached, one can either increase the accretion rate
or increase the value of the critical fastness. The latter option is illustrated
by the dotted curve, for which $\omega\sub{cr}=0.7$, twice the standard value
(see Section~\ref{fast}).

Once accretion stops, pulsars will start spinning down, so many pulsars
that lie to the right of any track can be explained by evolution along
that track down to some field strength and subsequent spindown. For
pulsars with ages greater than the age of the universe (i.e.\ below the
Hubble line) this will not work, so they must have emerged from the
accretion phase with long periods, close to the current values. This poses
no problems because for any evolution track that passes to the left of a
given pulsar there is one with a lower accretion rate and otherwise the
same parameters that passes through it. Such a low accretion rate may
impose constraints on binary evolution scenarios, but I shall not attempt
to exploit this here, and use only the requirement that the fastest
pulsars be produced (Section~\ref{fast}).

   \section{Constraints on $\bbeta$ and $\bmath{M\sub{c}}$}
   \label{bdmcon}

      \subsection{The weakest fields: PSR~J2229+2643 and J2317+1439}
      \label{weakest}

The two lowest-field pulsars known are PSR~J2229+2643 and J2317+1439
(Camilo et~al.\ 1996a)\nocite{cnt:96}. Both
are in binaries with low-mass white dwarf companions for which Van den
Heuvel and Bitzaraki (1995)\nocite{hb:95} estimated the amount of mass accreted
by the pulsar to be 0.70\msun\ and 0.79\msun, respectively. Some properties
are listed in Table~\ref{ta:systems}.
\begin{table*}
  \begin{minipage}{160mm}
  \caption[]{Relevant properties of the systems used. Radio pulsars are at
	   the top, X-ray pulsars at the bottom. $B\sub{m}$ is the field
	   measured from the radio pulsar spindown rate, $B\sub{e}$ is the
	   field after correcting for proper motion; for X-ray pulsars it
	   is the field for which the pulsar would be on the spinup line.
	   Pulsar companion masses are taken from Van den Heuvel \&
	   Bitzaraki (1995)\nocite{hb:95}. X-ray
	   binary parameters are taken from Van Kerkwijk, Van Paradijs \&
	   Zuiderwijk (1995)\nocite{kpz:95}.  $\me$ is estimated from the
	   X-ray luminosity.}
  \label{ta:systems} \newcommand{\twe}{\multicolumn{2}{c}{}}
  \begin{tabular}{lr@{.}lr@{.}lcr@{.}lrrlrl}\hline 
  name       & \multicolumn{2}{c}{$P\sub{spin}$}
	      & \multicolumn{2}{c}{$P\sub{orb}$}
		& \multicolumn{3}{c}{companion}
		& $\log B\sub{m}$ & $\log B\sub{e}$ & $\dot{M}_{-8}$ 
		   & $\tau\sub{X}$ & $\Delta M$ \\
	   & \multicolumn{2}{c}{(s)} & \multicolumn{2}{c}{(d)} 
	   &  type       &\multicolumn{2}{c}{mass (\msun)} & (G)  &  (G) 
	   &  & ($10^4$\,yr)& (\msun)\\ \hline
PSR\,J0218$+$4232 & 0&00232  &  2&03 &  He WD      &    0&20 & 8.63 &      &&&\\ 
PSR\,J1022$+$1001 & 0&0165   &  7&8  &  CO WD      & $>$0&45 & 8.92 &  8.8 &&& 0.01--0.045\\ 
PSR\,B1820$-$30A  &  0&00544 &  \twe &  single     &    \twe & 9.64 & \\ 
PSR\,B1821$-$24   &  0&00305 &  \twe &  single     &    \twe & 9.35 & \\ 
PSR\,B1831$-$00   &  0&521   &  1&81 &  He WD?     &    0&20?&10.94 &      &&& 0.80\\
PSR\,B1937$+$21   &  0&00155 &  \twe &  single     &    \twe & 8.61 &      &&&\\ 
PSR\,B1957$+$20   &  0&00161 &  0&382&  ?          & $>$0&02 & 8.22 &      &&&\\ 
PSR\,J2145$-$0750 & 0&0161   &  6&84 &  CO WD      & $>$0&45 & 8.84 &  8.6 &&& 0.01--0.045\\ 
PSR\,J2229$+$2643 &  0&00298 & 93&0  &  He WD      &    0&30 & 7.88 &  7.6 &&& 0.70\\
PSR\,J2317$+$1439 &  0&00345 &  2&46 &  He WD      &    0&21 & 7.97 &  7.8 &&& 0.79\\
~\\ 
SMC\,X-1     & 0&717   &  3&89 & B0Ib        &   15&   &      & 12.0
					& 4 & 2 & $8\times10^{-4}$ \\
LMC\,X-4     & 13&51    &  1&41 & O7III--V    &   16&   &      & 13.6
					& 5 & 1 & $5\times10^{-4}$ \\
Cen\,X-3     &  4&84    &  2&09 & O6.5II--III &   19&   &      & 12.7
					& 1 & 1 & $1\times10^{-4}$ \\
Her\,X-1     &  1&24    &  1&7  & A--FIV      &    2&3  &      & 11.7
					&0.2& 30& $6\times10^{-4}$ \\ \hline
  \end{tabular} 
  \end{minipage} 
\end{table*}

\begin{figure}
   \epsfxsize=84mm\epsfbox{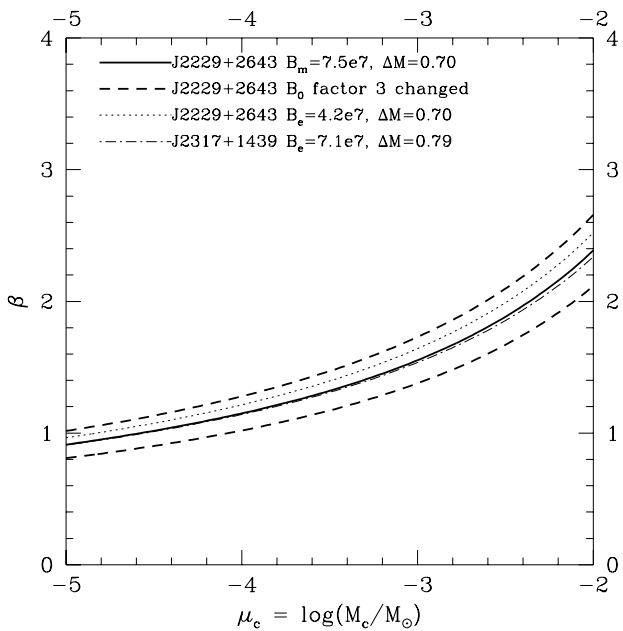} 
   \caption{Constraints in the $(M\sub{c},\beta)$ plane imposed by the
	    lowest-field pulsars}
   \label{fi:lowb} 
\end{figure}

For PSR\,J2229$+$2643, I use the measured field $B\sub{m}$ because the
proper motion has not been measured and therefore its $B\sub{e}$ is only
based on a typical value (see above). For PSR\,J2317$+$1439 the proper
motion is measured, so I use $B\sub{e}$. The solid line in
Fig.~\ref{fi:lowb} is the locus of parameter values that yields the field
of PSR\,J2229$+$2643 for its $\Delta M$, and the dashed-dotted curve the
same for PSR\,J2317$+$1439, both for $\log B_0=12.3$.
 They are almost the same, since the slightly lower field of
 PSR\,J2317$+$1439 and its higher $\Delta M$ act in opposite directions.
Using $B\sub{e}$ for PSR\,J2229$+$2643 gives the dotted curve.  The higher
curve here gives the strongest overall constraint, so I use the solid
curve in our final assembly of constraints, but I still have to account
for the variation in $B_0$. The dashed curves are as the solid one, but
for $\log B_0 =$ 12.8 (upper) and 11.8 (lower).  They also roughly bracket
the range 0.4--1.5\msun\ for $\Delta M$ at fixed $\log B_0=12.3$.  Given
the uncertainty in $B_0$, the entire region between the two dashed curves
satisfies the requirement that low enough fields be attainable.

      \subsection{The fastest pulsars: PSR\,B1937$\bmath+$21 
                  and B1821$\bmath-$24}
      \label{fast}

\begin{figure}
   \epsfxsize=84mm\epsfbox{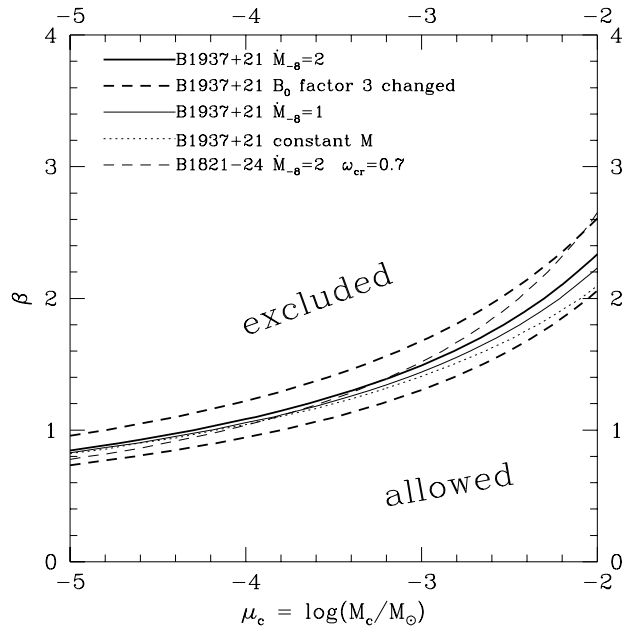}
   \caption{Constraints in the $(M\sub{c},\beta)$ plane imposed by the 
	    fastest pulsars.}
   \label{fi:fast}
\end{figure}

As explained above (Section~\ref{examp}), the only spin constraint applied
is that fast enough pulsars can be formed.  For $\me=2$ and
$\omega\sub{cr}=0.35$ the strongest constraint on the model is set by
PSR\,B1937$+$21, the lowest-field pulsar close to the spinup line. The
thick solid line in Fig.~\ref{fi:fast} indicates the locus of models that
pass exactly through its current position in the period-field diagram, for
$\log B_0=12.3$.  All models below the curve are allowed, those above it
are excluded.  The thick dashed lines indicate how the curve changes for
$\log B_0=$ 12.8 (upper) and 11.8 (lower).  The next strongest constraints
for the same accretion rate are provided by PSR\,B1957$+$20 and
J0218$+$4232.  Their curves for $\log B_0=12.3$ nearly coincide with the
upper dashed line and have not been plotted separately. For no curve shown
is the required amount of accreted mass greater than 0.8\msun, and for the
thick solid line it is about 0.3\msun\ everywhere on it.  The constraint
is tighter for a lower accretion rate because it puts PSR\,B1937$+$21
closer to the equilibrium period. The thin solid curve differs from the
thick one only in that $\me=1$, but the difference is small. The dotted
curve illustrates what happens if the change of mass of the neutron star
is neglected, but all parameters are as for the thick solid curve: it
leads to an artificially stronger constraint because it does not allow the
spinup line to move to shorter periods as the neutron star mass increases
(see Fig.~\ref{fi:bp}).

Next we must consider what to do with PSR\,B1821$-$24 and B1820$-$30A,
which are above the standard spinup line. 
Both these pulsars are in globular clusters and it is possible that their
location is a consequence of a type of evolution that is unique to that
environment (e.g.\ a dramatic period of mass transfer following a collision
with another star), so perhaps we should omit them from the discussion.
The location of PSR\,B1820$-$30A could even be a direct result of 
contamination of its period derivative due to acceleration in the
cluster potential (Biggs et~al.\ 1994)\nocite{bmlgf:94}.
But since I use standard spinup theory,
it is appropriate to at least briefly ask whether their 
location above the standard spinup line indicates a flaw in that theory that 
is serious enough to reconsider its use. 

Increasing the mass of the neutron star helps to bring it above the
standard spinup line, but it turns out that models in which these still
relatively high-field neutron stars have become heavy enough cannot produce
the lowest observed fields. One possible remedy is to accept higher
accretion rates, as observed in some X-ray pulsars (see Sect.~\ref{xray}).
If we set $\me=6.2$ both pulsars are on the spinup line. It turns out that
the strongest constraint on models is then set by PSR\,B1821$-$24, and
nearly coincides with the lower dashed line, so it would provide a stronger
constraint. Another option is to increase the critical fastness to 0.7 by
using a toy model
$n(\omega\sub{s})=1.39(1-\omega\sub{s}/0.7)/(1-\omega\sub{s})$. Once again
PSR\,B1821$-$24 provides the strongest constraint, shown by the thin dashed
curve ($\log B_0=12.3$); varying the initial field produces similar offsets
to the case of PSR\,B1937$+$21. In other words, simple and reasonable
adjustments to standard spinup theory can accommodate these two pulsars, and
would give stronger constraints. Due to the location of the pulsars in
globular clusters and the consequent possibility that they evolved in a
manner radically different from disc pulsars, it is better not to use those
constraints.

The weakest constraint from pulsar fastness is the upper dashed curve or
the thin dashed curve displaced upwards as appropriate for $\log
B_0=12.8$. Which one is used matters little once it is combined with the
constraint set in the previous section. The area that satisfies the most
conservative constraints from Figs.~\ref{fi:lowb} and \ref{fi:fast} is
shown in Fig.~\ref{fi:summ} as the hatched region. It is mostly determined
by the field of PSR\,J2229+2643, with a minor contribution from the spin
of PSR\,B1937$+$21 at low $\log M\sub{c}$.

      \subsection{X-ray pulsars}
      \label{xray}

The classical high-mass X-ray binaries contain high-field neutron stars.
To estimate their amounts of accreted mass we need the accretion rate,
which can be derived from the observed X-ray luminosity, and the X-ray
life time, for which models are required. The finally adopted values for
the four X-ray binaries SMC\,X-1, LMC\,X-4, Cen\,X-3, and Her\,X-1 are
listed in Table~\ref{ta:systems}.

The X-ray luminosities were taken from White, Swank, \& Holt 
(1983)\nocite{wsh:83}
and are somewhat higher than often quoted. This is because X-ray pulsars
have hard spectra and usual fluxes are cited in the range 2--10\,keV.
But White \etal\ quote luminosities from 0.5--60\,keV, which much better
represent the total luminosity, as can be seen from the spectra they
show. All four sources are disc accretors and are not expected to vary too
much in flux. For example, a low flux is sometimes seen for SMC\,X-1
%
%
(e.g.\ Seward \& Mitchell 1981\nocite{sm:81}),
 but a 7-year monitoring using the Vela 5B satellite
%
%
(Whitlock \& Lockner 1994)\nocite{wl2:94} shows an average 3--12\,keV
luminosity of $2\times10^{38}\un{erg}{}\un{s}{-1}$ despite the occasional
low state, quite consistent with the 2.5 times higher value for the wider
energy range quoted by White \etal\ The X-ray luminosities were converted
to accretion rates assuming that $L\sub{X} = GM\dot{M}/R$, with
$M=1.4\msun$ and $R=10$\,km.

One might worry that beaming of the X-ray pulsar radiation causes an
overestimate the luminosity because we are hit by the beam but other
positions in the sky are not, and our estimates assume isotropic emission.
For radio pulsars, which have very narrow beams, this clearly a concern.
But X-ray pulsar profiles are very different: inspection of the pulse profiles
in White et~al.\ (1983) reveals that significant emission is present over
the full 360\degs\ of pulse phase, and the rms.\ variation of the intensity
over phase is 0\% for Cen\,X-4, 5--20\% for SMC\,X-1, 20--50\% for Her\,X-1
and 50\% for Cen\,X-3. (Ranges are given if the modulation depth depends on
energy band.) Since these are very bright sources, they would be detected
even if the phase-average flux were as low as the lowest flux that occurs
at any phase.  This is unlike radio pulsars, many of which go undetected
because their beam is never pointed towards us.  Since the flux variation
in the direction perpendicular to the beam sweep is unlikely to be very
different from the one along the pulse, X-ray pulsars are detectable from
any direction in the sky. This means that (i) the phase-averaged flux we
see is unlikely to deviate much from the mean over the entire sphere
around the pulsar and (ii) the deviation could be positive or negative,
i.e.\ in a sample of pulsars we will not systematically overestimate the
luminosity.  Therefore, the estimated amounts of transferred mass are not
systematically biased upwards due to beaming.

To estimate the X-ray life times I use the model of beginning atmospheric
Roche lobe overflow (Savonije 1978, 1979)\nocite{savon:78,savon:79}. 
In this model,
the donor star is still burning hydrogen in its core, but extended in size
due to mass loss. It expands on a nuclear time scale and the mass transfer
rate grows exponentially with an e-folding time, $\tau\sub{X}$,
equal to the time it takes
the star to expand by one scale height. For massive donors this time scale
is about $10^4$\,yr, in agreement with the estimated X-ray life time
from population statistics (Meurs \& Van den Heuvel 1989\nocite{mh:89}).
Her\,X-1 has a subgiant donor of much lower mass, so its life time is
rather longer.  The values shown in Table~\ref{ta:systems} are taken from
Savonije. Since the accretion rate grows exponentially, $\Delta M$ is well
approximated by $\tau\sub{X}$ times the current accretion rate.

The magnetic fields of X-ray pulsars are high, but precise values are hard
to specify. The estimate in Table~\ref{ta:systems} is based on the
assumption that the pulsar is on the equilibrium spinup line, but only for
Her\,X-1 is the spinup time scale sufficiently shorter than $\tau\sub{X}$
to justify this assumption.  The cyclotron line feature seen in the
spectrum of \mbox{Her\,X-1} indicates a surface field of $3\times10^{12}$\,G
(Tr\"umper \etal\ 1978)\nocite{tprvs:78}, in good agreement with the spinup
estimate given that the cyclotron line probes the field very near the
neutron star, whereas the spinup estimate probes the large-scale dipole
field. No cyclotron lines have been seen from the other three X-ray
pulsars.  Since they had lower accretion rates and therefore longer
equilibrium periods in the past and are spinning up, their equilibrium
periods are probably shorter and therefore the field strengths are
overestimates.

\begin{figure}
   \epsfxsize=84mm\epsfbox{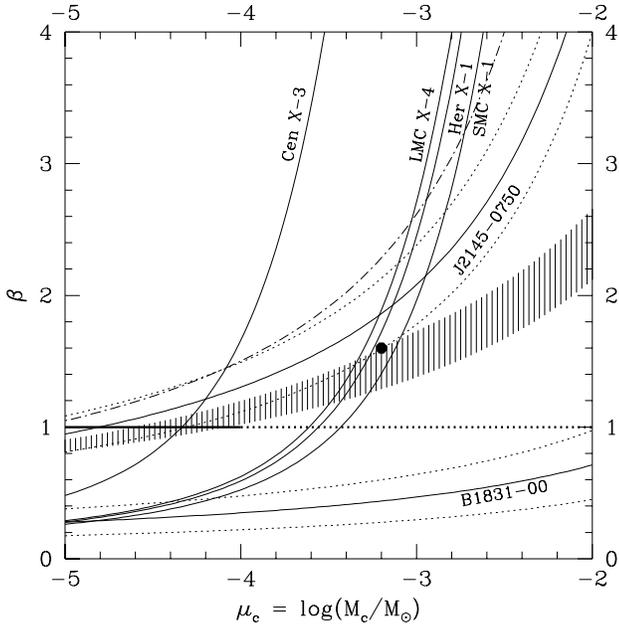}
   \caption{Summary of all constraints in the $(M\sub{c},\beta)$ plane.
	    The black dot marks the location of the model that is still
	    marginally consistent with all constraints except PSR\,B1831$-$00.}
   \label{fi:summ}
\end{figure}

If we assume that the four pulsars have had a factor 3 field decrease each
one defines a curve in the $(M\sub{c},\beta)$ plane that is shown in
Fig.~\ref{fi:summ}.  (A small systematic difference between their fields
and those of standard radio pulsars cannot be ruled out, but a factor 10 is
too much. A factor 3 is a reasonable estimate of what might still be
allowed.) The area below and to the right of the curves corresponds to lower
amounts of field decrease and is thus allowed. The area above them is
excluded. The three curves defined by LMC\,X-4, Her\,X-1, and SMC\,X-1 are
quite similar.  I shall take the middle one
(Her\,X-1) to fairly represent the X-ray pulsar constraint. The allowed
region thus left no longer contains any combination with $\beta=1$, so the
standard model in which field decay is proportional to accreted mass is
already ruled out.

      \subsection{PSR\,J2145$\bmath-$0750}
      \label{coup}

A recently discovered class of millisecond pulsars with high-mass CO white
dwarf companions provides us with a probe of the regime of $\Delta M$ in
between the X-ray pulsars and the most strongly recycled pulsars. Van den
Heuvel (1994)\nocite{heuve2:94} discusses their evolution in detail and
concludes that the progenitor of the white dwarf was a 1--6\msun\ star that
came into Roche contact on the asymptotic giant branch.  The amount of
accreted mass is mainly from wind accretion and increases from 0.01 to
0.045\msun\ with increasing donor mass. For the highest, least constraining
amount of accreted mass, and for $\log B\sub{m}=8.84$ for PSR\,J2145$-$0750
(based on $\dot{P}=2.98\times10^{-20}$, F. Camilo, private communication)
the curve of acceptable parameter pairs is shown in \Fig{fi:summ} for $\log
B_0=12.3$ (solid), 12.8 (upper dotted), and 11.8 (lower dotted). The lower
curve just grazes the still acceptable region at $(-3.2,1.6)$, so if one
wants to keep a field decay model from this class these are its parameters.
More likely, one cannot in reason stretch all the constraints to their
limits simultaneously and the model is inconsistent with the whole body of
available data. The best prospect for ruling it out altogether comes from
PSR\,J2145$-$0750 and its twin sister, PSR\,J1022$+$1001 (Camilo \etal\
1996a)\nocite{cnst:96}.  Both have significant estimated proper-motion
corrections to the magnetic field (see Fig.~\ref{fi:bp} and
Table~\ref{ta:systems}), so the lower values of the corrected fields will
push the curve up and remove it further from the area allowed by the
constraints. Also, not all will have had the maximum allowed accreted mass.
To indicate the strong effect of this I show the acceptable curve for
PSR\,J2145$-$0750 for the same parameters as the solid curve except that
$\Delta M=0.02\msun$ (dashed-dotted). It leaves no acceptable parameter
combinations.

   \section{Discussion}
   \label{discu}

      \subsection{PSR\,B1831$-$00 and 4U\,1626$-$67: Accretion-induced 
		  collapse?}
      \label{discu.aic}

In the previous section I have deliberately not included two objects that
have been adduced as good evidence against field decay proportional to
accreted mass in the past: PSR\,B1831$-$00 and the X-ray pulsar 4U\,1626$-$67.
The ages and orbital parameters of both suggest that the neutron star has
accreted at least 0.5\msun, yet the fields are high (Verbunt \etal\ 
1990)\nocite{vwb:90}.  The
reason is that at various times accretion-induced collapse (AIC) of a white
dwarf has been proposed as a method of making these, allowing the neutron
star to have accreted less mass than it seems, so I wanted to make the
case against the decay model without them. Also, both have peculiarly low
companion masses so they may really be different.

But let us briefly revisit PSR\,B1831$-$00. It has $B=8.7\times10^{10}$\,G and
$P\sub{orb}=1.8$\,d. Van den Heuvel and Bitzaraki (1995)\nocite{hb:95} claim
that its position at high field is good evidence for accretion-induced
collapse of a white dwarf. In this scenario, the accreting object in the
binary was a white dwarf while most of the donor mass was accreted, but
near the end of accretion it reached the Chandrasekhar mass and collapsed
to a neutron star, which then accreted little enough to keep a high field.
For the standard $\beta=1$ model they used this may still be reasonable,
since the accreted amount after collapse could be 0.1\msun. This is a fair
fraction of the total transferred ($\sim0.8\msun$) so the moment of
collapse need not be tuned too delicately. But the most tolerable model
still left now ($\beta=1.6,\muc=-3.2$) limits the amount of accreted mass
to 0.01\msun, even if we allow the pulsar to start with a
$10^{13}$\,G field. This does require the collapse to be rather finely
timed very near the end of accretion, and makes the scenario quite
unattractive. Even if one were willing to accept this fine tuning for
PSR\,B1831$-$00, there is still the following coincidence: why should it
be the only short-period binary pulsar with such a high field? After all,
there must have been less fortunately timed cases like it, in which the
collapse occurred less close to the end of mass transfer. The resulting
neutron stars would have lower fields, between that of PSR\,B1831$-$00 and
the canonical low values, and therefore live longer and be more abundant.
Note that this problem occurs no matter which parameters one chooses for
the decay model. Consequently, I feel that one should accept the fact that
it has accreted significant amounts of mass, as has 4U\,1626$-$67.  This is
of course fatal to the decay model, as illustrated by the lowest curves in
Fig.~\ref{fi:summ}, which define the region of acceptable parameters to fit
PSR\,B1831$-$00 (using $\Delta M=0.8\msun$ and $\log B_0=11.8, 12.3,
12.8$).

AIC has at various times in the past been called on to rescue the then
popular field decay model, by providing an alternative formation mechanism
for the few objects that obviously failed to fit the model. So it was when
\mbox{Her\,X-1} failed to fit the rapid exponential decay of neutron star
magnetic fields, but Verbunt et~al.\ (1990) showed that AIC cannot provide
a viable alternative in that case and that therefore Her\,X-1 does disprove
the model. It will remain a matter of taste in each case whether to view
the lone strange object as an exception that proves the model or a nail its
coffin. But in the case of PSR\,B1831$-$00 and 4U\,1626$-$67 the exceptions
are once again nails, if not the only ones, and AIC cannot change this
(This argument was already partly given by Verbunt \etal\ 1990.)

      \subsection{Constraints on field only}
      \label{discu.field}

\begin{figure}
   \epsfxsize=84mm\epsfbox{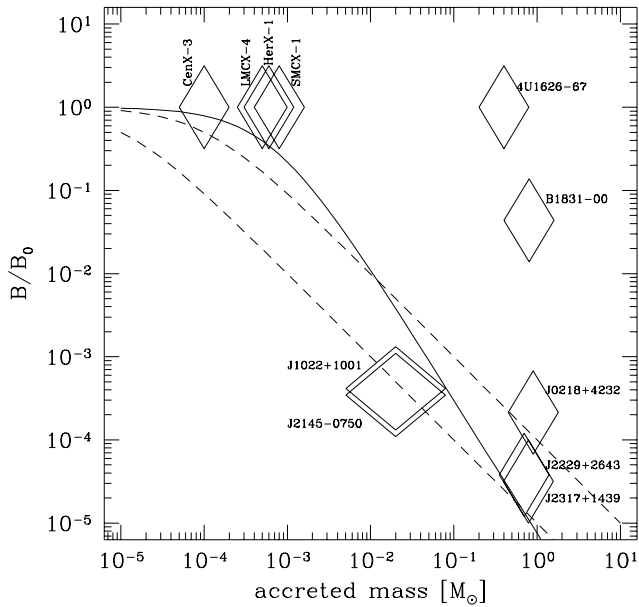}
   \caption{The accreted mass versus magnetic field for all objects in
	    Table~\ref{ta:systems} except those that constrain the model
	    through their spin period. The dashed curves are decay models
	    with $\beta=1$ and $\muc=-4$ (top) and $-5$ (bottom). The
	    solid curve is for the model marked with a black dot in
	    \Fig{fi:summ} ($\beta=-1.6,\muc=-3.2$).
            \label{fi:bdm}
	    }
\end{figure}

Since the spin period constraints did not contribute too much
to constraining the field decay model, it is possible to capture most of 
the argument in a straight plot of amount of field change versus 
accreted mass (\Fig{fi:bdm}). Each diamond represents one object whose
name is plotted alongside it. The size of the diamond 
represents the error in the location of the object. Most of the `errors'
in the quantities ultimately include uncertainties in the models used
to interpret the data from which the quantities are derived and they 
therefore are indicative of the author's personal assessment of those
models rather than well-defined statistical quantities. The error in the ratio
$B/B_0$ was taken to be 0.5 dex and mainly represents the variation of
initial magnetic fields of neutron stars; this range nominally covers
87\% of initial fields (\Sect{initcon}). The amounts of accreted mass 
for the X-ray pulsars are derived partly from theory and partly from data.
Those of the millisecond pulsars are related via theory to system
parameters that are accurately known observationally, so the error is
mostly uncertainty in the theory. In all cases, I have assigned an error
of a factor 2 either way to the estimated $\Delta M$, which should be fairly
generous. For PSR\,J1022$+$1001 and J2145$-$0750, the theoretical
estimate itself already spans a factor 2 each way from the mean of 
0.02\msun\ (Table~\ref{ta:systems}), so I have used a total error of
a factor 4 either way for their $\Delta M$.
Given the generous errors used one should really demand that any model
curve pass through 80\% of all the diamonds. For any diamond that is
widely missed, one will have to find an alternative model that puts the
object in a completely different position.

The first striking impression from \Fig{fi:bdm} is that the full range of
accreted masses is represented by objects with little or no field decay, and
the full range of amounts of field decay is represented by objects that
have accreted around 0.5\msun. This bodes ill for any model
that demands a strict relation between the two quantities.  A few models
are indicated as well: the dashed curves represent the standard version of
the model, $\beta=1$. They lie either too low for the X-ray pulsars or too
high for PSR\,J1022$+$1001 and J2145$-$0750.  The solid curve represents
the marginal model indicated by the black dot in \Fig{fi:summ}. It just
grazes the corners of many error boxes, and is already in trouble at the
right of the diagram, since its predicted fields there are almost too low
for the lowest-field pulsars known, but many pulsars exist at that same 
amount of accreted mass and ten times higher fields (like 
PSR\,J0218$+$4232).

The really difficult objects for the accretion-decay model are 
4U\,1626$-$67 and PSR\,B1831$-$00. They cannot possibly satisfy any
model of this type. The difficulties of advancing accretion-induced collapse
for the formation of PSR\,B1831$-$00 (\sect{discu.aic}) are even greater
for 4U\,1626$-$67, where the amount of accreted mass needs to be reduced
by 3 orders of magnitude from the total amount of mass transfer that has
occurred in the system before it comes close to fitting any of the models.
But mass transfer in this system is still ongoing and very long-lived, so
this neutron star would have to have formed almost literally yesterday by AIC
to save the accretion-decay model.

It is tempting to conclude from \Fig{fi:bdm} that a significant amount
of accreted mass, while it does not guarantee much field decay, is at least
a necessary condition for it, because there appear to be no objects in the
lower left corner. But that conclusion is not valid because this 
absence may well be due to poor chances of detecting
such sources. Imagine a neutron star with very little accreted mass that
has had much field decay. If it were not accreting it could only be visible
as a radio pulsar. But for it to be above the death line it would have
to have been spun up, which is not possible without significant mass accretion.
If it were an accreting source, chances of detection are small again because
the small amount of accreted mass implies that either the accretion phase
has to be very short or the accretion rate has to be very low.

      \subsection{Further implications for models}
      \label{discu.furt}

The detailed model by Romani (1993)\nocite{roman:93} does predict field
decay proportional to accreted mass down to a bottom value of around
$10^8$\,G with $M\sub{c} = 10^{-5}-10^{-4}\msun$, as long as the accretion
rate is high. Since all the objects used in the present paper have accreted
most of their mass at rates not far below the Eddington rate this model as
it stands is excluded by the data.

The boundaries I have drawn are statistical in nature due to the possible
variations in initial magnetic field (\Sect{initcon}). But the range used
everywhere covers almost 90\% of initial fields, and for every type of
constraint there are at least 2 objects which provide a nearly equally
strong version of that constraint. Therefore it is highly unlikely that we
have condemned the model unjustly because all pulsars used to constrain it
are untypical of the average in a direction that is unlucky for the
model.

It is of course possible to save the model by varying the parameters
$\beta$ and $M\sub{c}$ between pulsars, but this would largely remove its
attraction unless one had a prescription for what parameters
to give to which kind of pulsar. Doing so would merely reinforce the
main conclusion of the present paper, namely that other parameters than
accreted mass alone influence significantly the decay of a neutron star's
magnetic field.

   \section{Conclusion}
   \label{conclu}

I have explored the consequences of a popular neutron star magnetic-field
decay model in which the decay is determined solely by the amount of
accreted mass.  The result is that no model of this type (Eq.~\ref{eq:bm})
can satisfy all the known constraints derived from the fields, estimated
amounts of accreted mass, and spin periods of a variety of recycled radio
pulsars and X-ray binaries.  Only if the constraints are all stretched to
their limit simultaneously and PSR\,B1831$-$00 and 4U\,1626$-$67 are not
required to fit is there a small parameter range for which it is still
viable.  The most used version of the model, in which field decay is
proportional to the amount of accreted mass to the first power, is firmly
ruled out even then.

An important constraint is provided by the pulsars
with very high apparent ages near the Hubble line. Their observed period
derivatives are affected by proper motion and should be revised downwards
(Camilo \etal\ 1994)\nocite{ctk:94}. Once more of their proper motions have been measured the
constraints on recycling models set by them will strengthen significantly.
The other important constraint comes from the fact that some X-ray pulsars
have high fields despite having accreted up to $10^{-3}\msun$. Since this
value is somewhat model dependent, a firmer quantitative understanding of
the evolution of classical high-mass X-ray binaries would also help to
constrain models of neutron star field decay.

   \section*{Acknowledgements}

I would like to thank F. Camilo for helpful comments during the preparation
of this paper, the early communication of some of his results, and careful
reading of an early version of the manuscript, and J. van Paradijs
for his helpful suggestions.


\end{document}